\documentclass[aps,prb,twocolumn]{revtex4-1}

\usepackage[english]{babel}
\usepackage[utf8]{inputenc}
\usepackage{amsmath}
\usepackage{amssymb}
\usepackage[caption=false]{subfig}
\usepackage{amssymb}
\usepackage{epsfig}
\usepackage{graphicx}
\usepackage{amsmath}
\usepackage{array,color}
\usepackage{natbib}

\usepackage{soul}
\usepackage{textcomp}
\usepackage[T1]{fontenc}

\usepackage[usenames,dvipsnames]{xcolor}
\definecolor{forestgreen}{rgb}{0.11,0.54,0.15}
\definecolor{purple}{rgb}{0.62,0.10,0.96}
\definecolor{dockerblue}{rgb}{0.11,0.56,0.98}
\definecolor{freeblue}{rgb}{0.25,0.41,0.88}

\usepackage[pdftex,plainpages=false,colorlinks=true,linkcolor=Red, citecolor=blue, urlcolor=blue]{hyperref}

\usepackage{latexsym}
\usepackage{url}
\usepackage{xcolor}

\usepackage{bm}

\usepackage{etoolbox}

\begin{document}

\title{Towards non-parametric fiber-specific $T_1$ relaxometry in the human brain}

\author{A. Reymbaut$^{1,2}$}
\email{alexis.reymbaut@fkem1.lu.se}
\author{J. Critchley$^3$}
\author{G. Durighel$^3$}
\author{T. Sprenger$^{4,5}$}
\author{M. Sughrue$^6$}
\author{K. Bryskhe$^{2}$}
\author{D. Topgaard$^{1,2}$}

\affiliation{
$^1$Department of Physical Chemistry, Lund University, Lund, Sweden\\
$^2$Random Walk Imaging AB, Lund, Sweden\\
$^3$Spectrum Medical Imaging, Sydney, Australia\\
$^4$Karolinska Institute, Stockholm, Sweden\\
$^5$GE Healthcare, Stockholm, Sweden\\
$^6$Charlie Teo Foundation, Sydney, Australia
}

\date{\today}

\begin{abstract}
\textbf{Purpose:} To estimate fiber-specific $T_1$ values, \textit{i.e.} proxies for myelin content, in heterogeneous brain tissue.\\
\textbf{Methods:} A diffusion-$T_1$ correlation experiment was carried out on an \textit{in vivo} human brain using tensor-valued diffusion encoding and multiple repetition times. The acquired data was inverted using a Monte-Carlo inversion algorithm that retrieves non-parametric distributions $\mathcal{P}(\mathbf{D},R_1)$ of diffusion tensors and longitudinal relaxation rates $R_1 = 1/T_1$. Orientation distribution functions (ODFs) of the highly anisotropic components of $\mathcal{P}(\mathbf{D},R_1)$ were defined to visualize orientation-specific diffusion-relaxation properties. Finally, Monte-Carlo density-peak clustering (MC-DPC) was performed to quantify fiber-specific features and investigate microstructural differences between white-matter fiber bundles. \\
\textbf{Results:} Parameter maps corresponding to $\mathcal{P}(\mathbf{D},R_1)$'s statistical descriptors were obtained, exhibiting the expected $R_1$ contrast between brain-tissue types. Our ODFs recovered local orientations consistent with the known anatomy and indicated possible differences in $T_1$ relaxation between major fiber bundles. These differences, confirmed by MC-DPC, were in qualitative agreement with previous model-based works but seem biased by the limitations of our current experimental setup. \\
\textbf{Conclusions:} Our Monte-Carlo framework enables the non-parametric estimation of fiber-specific diffusion-$T_1$ features, thereby showing potential for characterizing developmental or pathological changes in $T_1$ within a given fiber bundle, and for investigating inter-bundle $T_1$ differences. 
\end{abstract}

\maketitle

\footnotetext[1]{\textbf{Abbreviations used:} MRI, magnetic resonance imaging; drMRI, diffusion-relaxation MRI; ODF, orientation distribution function; MC-DPC, Monte-Carlo density-peak clustering; DPC, density-peak clustering; CSF, cerebrospinal fluid; GM, grey matter; WM, white matter; CC, corpus callosum; CING, cingulum; AF, arcuate fasciculus; CST, corticospinal tract.}

\newpage

\section{Introduction}
\label{Sec_Intro}

While diffusion MRI has provided enhanced sensitivity to tissue microstructure \textit{in vivo} by capturing the translational motion of water molecules diffusing in biological tissue,~\citep{LeBihan:1990, LeBihan:1992, Basser:1994, Jones:2010} diffusion-relaxation MRI (drMRI) additionally reports on the local chemical composition of the aqueous phase.~\citep{Zhang:2014,deSantis_relaxometry:2016, deSantis_T1:2016, Hutter:2018,Park_ISMRM:2018,Slator:2019,deAlmeidaMartins_Topgaard:2018,deAlmeidaMartins:2020} For instance, the longitudinal relaxation time $T_1$ informs on molecular dynamics~\citep{Bloembergen:1947} and on the presence of paramagnetic species~\citep{Bloembergen:1948} in simple liquids. \textit{In vivo}, it is mainly determined by cross relaxation, magnetization transfer and spin diffusion with macromolecules in general~\citep{Edzes_Samulski:1977, Halle:2006,Rooney:2007} and myelin lipids in particular,~\citep{Mottershead:2003, Bjarnason:2005, deSantis:2014, Lutti:2014, Stuber:2014} as well as by the interplay between relaxation and diffusion.~\citep{Brownstein_Tarr:1977} However, microstructural studies have been hindered by the fact that the measured drMRI signal is only sensitive to the voxel-averaged diffusion-relaxation profile, with typical cubic-millimeter voxels comprising multiple cell types and the extra-cellular space.~\citep{Stanisz:1997,Norris:2001,Sehy:2002,Minati:2007,Mulkern:2009} 

Two strategies were explored to alleviate the lack of specificity of the drMRI signal. 
On the one hand, multiple models and signal representations have been developed to relate either the diffusion-$T_2$~\citep{Veraart:2018,Lemberskiy:2018,Ning:2020} or diffusion-$T_1$~\citep{deSantis_relaxometry:2016, deSantis_T1:2016,Andrews_ISMRM:2019} MRI signal to the voxel content. However, models/signal representations rely on compartmental/functional assumptions that may disagree with the underlying tissue microstructure.~\citep{Jelescu:2017,Novikov_on_modeling:2018,Reymbaut_accuracy_precision:2020}
On the other hand, `tensor-valued' diffusion encoding gradient waveforms have enhanced the specificity of the data itself by targeting specific features of the intra-voxel diffusion profile.~\citep{Eriksson:2013,Westin:2014,Eriksson:2015,Westin:2016,Topgaard:2017,Topgaard_dim_rand_walks:2019} Indeed, such measurements typically use four acquisition dimensions - the trace $\mathit{b}$ (size), normalized anisotropy $\mathit{b}_\Delta\in [-0.5,1]$ (shape) and orientation $(\Theta,\Phi)$ of an axisymmetric encoding tensor $\mathbf{b}$~\citep{Basser:1994,Mattiello:1994,Mattiello:1997} - to probe the four dimensions of microscopic axisymmetric diffusion tensors, \textit{i.e.} their isotropic diffusivity $\mathit{D}_\mathrm{iso}$, normalized anisotropy $\mathit{D}_\Delta\in [-0.5,1]$ and orientation $(\theta,\phi)$.~\citep{Haeberlen:1976, Conturo:1996} Tensor-valued diffusion acquisition schemes have since been used to further investigate signal representations~\citep{Lasic:2014, Westin:2016, Cottaar:2020} and models.~\citep{Lampinen_CODIVIDE:2017, Coelho:2019, Coelho:2019_MICCAI, Reisert:2019, Lampinen:2020, Reymbaut_arxiv_Magic_DIAMOND:2020} 

While inversion of the diffusion-relaxation NMR signal is already common practice in the porous media field,\citep{Prange:2009, Galvosas:2010, Bernin_Topgaard:2013, Song_book_chapter:2017} the advent of `tensor-valued' diffusion-relaxation correlation measurements has resulted in the development of non-parametric Monte-Carlo signal inversion algorithms of the diffusion~\citep{deAlmeidaMartins_Topgaard:2016} and diffusion-$T_1$-$T_2$~\citep{deAlmeidaMartins_Topgaard:2018} MRI signals in porous media, and of the diffusion~\citep{Topgaard:2019} and diffusion-$T_2$~\citep{deAlmeidaMartins:2020} MRI signals in the \textit{in vivo} brain. 
Although noise-sensitive,~\citep{Reymbaut_accuracy_precision:2020} these algorithms do not rely on any compartmental/functional assumption regarding the voxel content, nor on constraints regarding data compression~\citep{Venkataramanan:2002} or the density of the acquisition sampling scheme.~\citep{Benjamini:2016, Kim:2017, Benjamini:2018, Benjamini:2020, Kim:2020} They also do not consider any regularization to guide the search for a suitable solution to the inverse problem.~\citep{Provencher:1982,Kroeker:1986,Whittall:1989,Mitchell:2012} 
Enhanced by methods aiming to visualize and quantify fiber-specific properties, even in fiber-crossing areas of the white matter (60 to 90\% of voxels in a typical whole-brain imaging experiment~\citep{Jeurissen:2013}), Monte-Carlo signal inversions have been shown to yield critical sensitivity and specificity to fiber-specific $T_2$ values.~\citep{deAlmeidaMartins_ODF:2020, deAlmeidaMartins_thesis:2020, Reymbaut_arxiv_MC_DPC:2020} However, this work has yet to be extended to fiber-specific $\mathit{T}_1$-values, which are of particular interest to evaluate changes in bundle-specific myelin contents,~\citep{Liu:2019} relevant to the study of neurodevelopment, plasticity, aging and neurological disorders.~\citep{van_den_Heuvel:2010,Caeyenberghs:2016,Mancini:2018} Indeed, $T_1$ contrast is sensitive to myelin,~\citep{Mottershead:2003, Bjarnason:2005, deSantis:2014, Lutti:2014, Stuber:2014} like many other contrasts~\citep{Laule:2007, Campbell:2018} such as $T_2$,~\citep{Mackay:1994, Beaulieu:1998, Gareau:2000, Webb:2003, Stanisz:2004, Laule:2006, Laule:2008} $T_2^*$~\citep{Hwang:2010,Lee:2012,Sati:2013} and magnetization transfer.~\citep{Gareau:2000, Schmierer:2008} Importantly, note that none of these contrasts are `specific' to myelin.~\citep{Campbell:2018} 

In this work, we leverage non-parametric distributions $\mathcal{P}(\mathbf{D},R_1)$ of diffusion tensors $\mathbf{D}$ and longitudinal relaxation rates $R_1 = 1/T_1$ obtained \textit{via} Monte-Carlo inversion of a diffusion-$T_1$ weighted \textit{in vivo} human-brain dataset to resolve sub-voxel diffusion-$R_1$ components. We first estimate parameter maps of these distributions' statistical descriptors and extract orientation-resolved $T_1$ values within the pool of highly anisotropic components output by the Monte-Carlo inversion algorithm. These $T_1$ values are then color-mapped onto non-parametric orientation distribution functions (ODFs)~\citep{deAlmeidaMartins_ODF:2020, deAlmeidaMartins_thesis:2020} and quantified in terms of median value and precision using orientational clusters obtained \textit{via} Monte-Carlo density-peak clustering (MC-DPC).~\citep{Reymbaut_arxiv_MC_DPC:2020} In particular, these novel tools enable to identify significant differences with respect to $T_1$ relaxation between major white-matter bundles. After describing how our \textit{in vivo} human-brain data was acquired in Section~\ref{Sec_in_vivo_data}, we lay down the theory underlying the Monte-Carlo signal inversion algorithm, the statistical descriptors of $\mathcal{P}(\mathbf{D},R_1)$, and our ODFs in Section~\ref{Sec_Monte_Carlo}, and detail the MC-DPC procedure in Section~\ref{Sec_MC_DPC}. We then present our results in Section~\ref{Sec_Results} and discuss them in Section~\ref{Sec_Discussion}, before concluding in Section~\ref{Sec_Conclusions}. 

We emphasize that this work is merely a proof of concept for non-parametric fiber-specific $T_1$ relaxometry. Firstly, vast improvements could be brought to our experimental setup, as detailed in Sections~\ref{Sec_in_vivo_data} and \ref{Sec_Discussion}. Second, better matching between the output of our Monte-Carlo framework and plausible white-matter tracts would be yielded upon integrating tractography~\citep{Mori:1999,Basser:2000,Morris:2008,Reisert:2011,Christiaens:2015,Neher:2017,Konopleva:2018,Poulin:2019} into our analysis pipeline.






\section{Methods}
\label{Sec_Methods}

\subsection{In vivo human-brain data}
\label{Sec_in_vivo_data}

\begin{figure*}[ht!]
\begin{center}
\includegraphics[width=30pc]{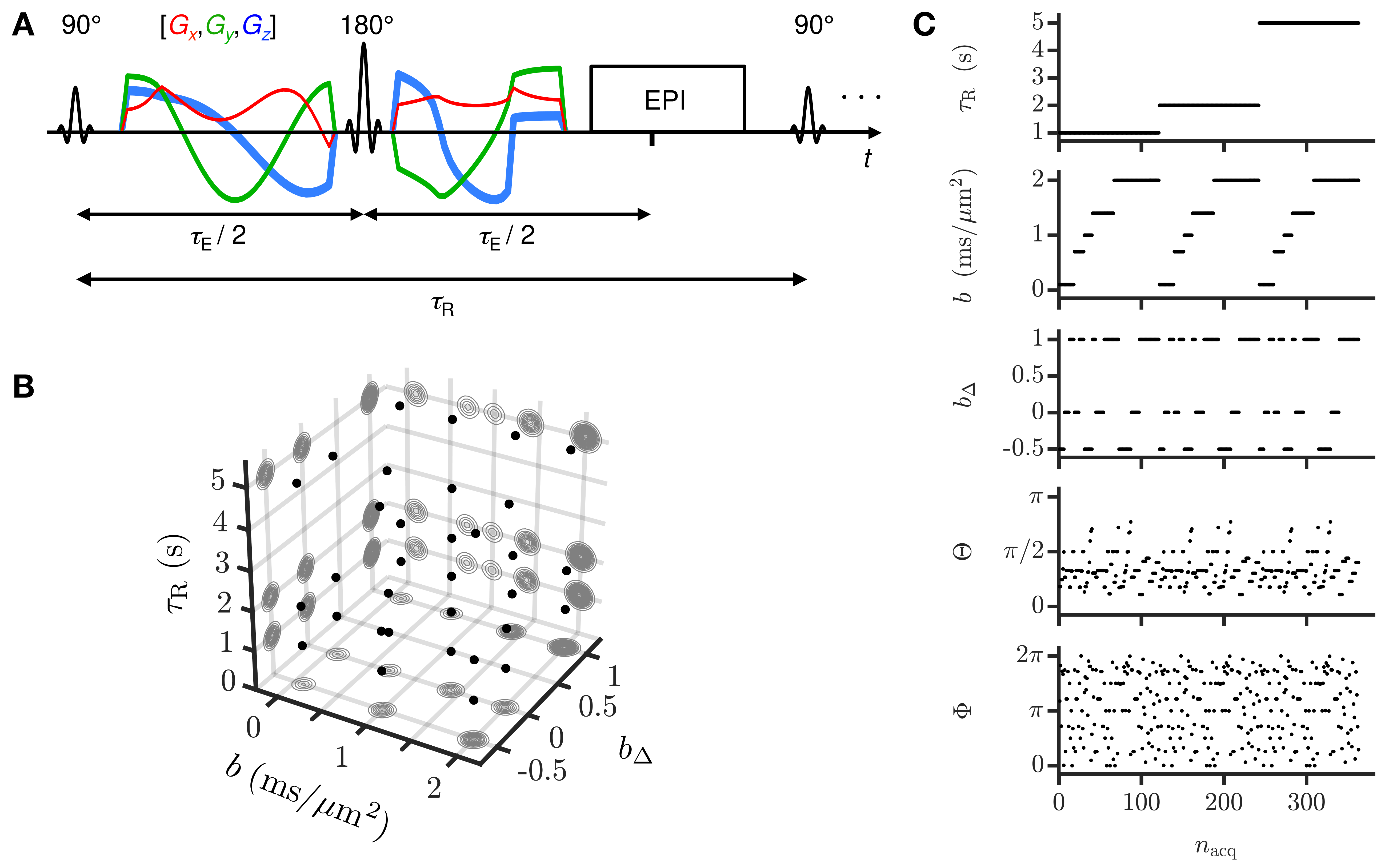}
\caption{Visualization of our acquisition scheme. (A) Spin-echo sequence with EPI readout customized for tensor-valued diffusion encoding and variable repetition time $\tau_\mathrm{R}$. (B) 5D grid-like acquisition scheme where black points indicate the acquisition points in the 3D subspace of repetition time $\tau_\mathrm{R}$, b-tensor size $b$ and b-tensor shape $b_\Delta \in [-0.5, 1]$. The number of b-tensor orientations $(\Theta,\Phi)$ for each point is illustrated by the projected contours. (C) Acquisition parameters as a function of sorted acquisition point index $n_\mathrm{acq}$.}
\label{Figure_acq}
\end{center}
\end{figure*}

Data collection was approved by the Spectrum Medical Imaging local ethics committee. A healthy volunteer was scanned on a 3T GE 750w equipped with a 32-channel receiver head and neck GEM coils (only 12-16 channels used for head) using a prototype GE multidimensional diffusion (MDD) spin-echo sequence with EPI readout, echo time $\tau_\mathrm{E} =120$ ms, FOV=240x240x12 mm$^3$, voxel-size=3x3x3 mm$^3$, fat-saturation (fat-sat) pulses,~\cite{Haase:1985} and ASSET acceleration factor=2, customized for tensor-valued diffusion encoding~\citep{Lasic:2014,Szczepankiewicz_DIVIDE:2019} and variable repetition time $\tau_\mathrm{R}$. Tensor-valued diffusion encoding was performed with numerically optimized~\citep{Sjolund:2015} Maxwell-compensated~\citep{Szczepankiewicz_Maxwell:2019} waveforms. We also attempted to match their frequency contents.~\citep{Lundell:2019} The same tensor-valued diffusion-weighted sequence was repeated for $\tau_\mathrm{R} =1$, 2 and 5 s. The dimensions of the resulting 20-minute 363-point acquisition scheme, shown in Figure~\ref{Figure_acq}, match those of $\mathcal{P}(\mathbf{D},R_1)$. The signal-to-noise ratio of this dataset was estimated across voxels of the corona radiata by computing the mean-to-standard-deviation ratio of the spherically encoded diffusion signal at $b=0.1$ ms/{\textmu}m$^2$ (see Supplemental Material of Ref.~\onlinecite{Szczepankiewicz_DIVIDE:2019}). It equals 20 at $\tau_\mathrm{R}=1$ s and 40 at $\tau_\mathrm{R}=5$ s.

As indicated by the aforementioned FOV and voxel size, only four axial slices were acquired so that to limit the acquisition time. While an inversion-recovery slice-shuffling sequence could drastically reduce acquisition time,~\citep{Hutter:2018, Park_ISMRM:2018} our prototype sequence is currently limited to sequential slices. 

\subsection{Non-parametric Monte-Carlo inversion}
\label{Sec_Monte_Carlo}

\subsubsection{Signal fitting and bootstrapping}
\label{Sec_MC_algorithm}


We used a modified version of the 5D Monte-Carlo inversion algorithm found in Ref.~\onlinecite{deAlmeidaMartins:2020} (pioneered by Ref.~\onlinecite{Prange:2009}) to analyze the diffusion-$R_1$ dataset described in Section~\ref{Sec_in_vivo_data}. Let us consider axisymmetric diffusion tensors, parametrized by their axial diffusivity $\mathit{D}_\parallel$, radial diffusivity $\mathit{D}_\perp$ and orientation $(\theta, \phi)$. An alternative parametrization includes the isotropic diffusivity $\mathit{D}_\mathrm{iso} = (\mathit{D}_\parallel + 2\mathit{D}_\perp)/3$ and normalized anisotropy $\mathit{D}_\Delta = (\mathit{D}_\parallel - \mathit{D}_\perp)/(\mathit{D}_\parallel + 2\mathit{D}_\perp) \in [-0.5, 1]$.~\citep{Haeberlen:1976, Conturo:1996, Eriksson:2015} Our Monte-Carlo inversion technique retrieves non-parametric 5D intra-voxel distributions $\mathcal{P}(\mathbf{D},\mathit{R}_1) \equiv \mathcal{P}(\mathit{D}_\parallel,\mathit{D}_\perp,\theta,\phi, \mathit{R}_1)$ by fitting a set of diffusion-$T_1$ weighted signals with a finite weighted sum of $\mathit{N}$ components $(\mathbf{D}_\mathit{n}, \mathit{R}_{1,\mathit{n}})\equiv (\mathit{D}_{\parallel,\mathit{n}}, \mathit{D}_{\perp,\mathit{n}} , \theta_\mathit{n}, \phi_\mathit{n}, \mathit{R}_{1,\mathit{n}})$, with $1\leq n\leq N=50$. Given that the $\mathit{T}_1$-weighting of the dataset detailed in Section~\ref{Sec_in_vivo_data} is provided through a spin-echo sequence with constant echo time $\tau_\mathrm{E}$ and variable repetition time $\tau_\mathrm{R}$, the inversion algorithm inverts the following discretized signal equation~\citep{Perman:1984}
\begin{align}
\mathcal{S}_\mathit{m} = & \sum_{\mathit{n}=1}^\mathit{N} \mathit{w}_\mathit{n}\,\exp(-\mathbf{b}_\mathit{m}:\mathbf{D}_\mathit{n}) \nonumber \\
 & \times \left[1 -2\exp([\tau_{\mathrm{E}}/2-\tau_{\mathrm{R},\mathit{m}}]\mathit{R}_{1,\mathit{n}}) + \exp(-\tau_{\mathrm{R},\mathit{m}}\mathit{R}_{1,\mathit{n}})\right]\, ,
\label{Eq_signal_discretized}
\end{align}
where $\mathcal{S}_\mathit{m}$ is the $m^\text{th}$ acquired signal, $\mathit{w}_\mathit{n}$ is the weight of the $n^\text{th}$ component, $\mathbf{b}$ is the diffusion-encoding tensor (b-tensor)~\citep{Mattiello:1994, Mattiello:1997} from tensor-valued diffusion encoding,~\citep{Eriksson:2013,Westin:2014,Eriksson:2015,Westin:2016,Topgaard:2017,Topgaard_dim_rand_walks:2019} and ":" is the Frobenius inner product. The weights $w_n$ are normalized so that $\sum_{\mathit{n}=1}^\mathit{N} \mathit{w}_\mathit{n} = \mathcal{S}_0 = \mathcal{S}(\mathbf{b}=\mathbf{0},\mathrm{\tau}_\mathrm{R}\to +\infty)$. For axisymmetric b-tensors, the Frobenius inner product writes~\citep{Eriksson:2015} $\mathbf{b}:\mathbf{D} = bD_\mathrm{iso}[1 + 2b_\Delta D_\Delta P_2(\cos\beta)]$, where $\mathit{P}_2(\mathit{x}) = (3\mathit{x}^2-1)/2$ is the second Legendre polynomial and $\cos\beta = \cos\Theta \cos\theta + \sin\Theta\sin\theta\cos(\Phi - \phi)$ is the cosine of the shortest angle $\beta$ between the main axis $(\Theta,\Phi)$ of $\mathbf{b}$ and the main axis $(\theta,\phi)$ of $\mathbf{D}$. 

A short-hand notation of Equation~\ref{Eq_signal_discretized} reads 
\begin{equation}
\mathbf{S} = \mathbf{K}\mathbf{w}\, ,
\end{equation}
where $\mathbf{S}$ is the column vector containing the acquired signals $\mathcal{S}_m$, $\mathbf{K}$ is the inversion kernel matrix containing the signal decays and $\mathbf{w}$ is the column vector containing the weights $\mathit{w}_n$ of the components $(\mathbf{D}_n, \mathit{R}_{1,n})$.
The Monte-Carlo inversion algorithm randomly samples such components within the following ranges, $D_\parallel,D_\perp \in [5\times 10^{-3}, 5]\;\text{\textmu}\mathrm{m}^2/\mathrm{ms}$, $\cos\theta\in[0,1[$, $\phi\in[0,2\pi[$ and $R_1\in[0.1,2]\;\mathrm{s}^{-1}$, and estimates the associated vector $\mathbf{w}$ quantifying the components' propensity to fit the acquired signals \textit{via} non-negative least-squares fitting:~\citep{Lawson_book:1974,Whittall:1989,English:1991,Venkataramanan:2002,Mitchell:2012}
\begin{equation}
\mathbf{w} = \underset{\mathbf{w}^\prime\geq 0}{\mathrm{argmin}}\;\Vert\mathbf{S}-\mathbf{K}\mathbf{w}^\prime\Vert_2^2\, ,
\end{equation}
where $\Vert\cdot\Vert_2$ denotes the L$_2$ norm. This process is repeated iteratively following a quasi-genetic filtering detailed in Refs.~\onlinecite{deAlmeidaMartins_Topgaard:2016, deAlmeidaMartins_Topgaard:2018, Topgaard:2019, deAlmeidaMartins:2020}.
%
Embracing the inherent ill-conditioning of Laplace inversion problems, we performed bootstrapping with replacement~\citep{de_Kort:2014} on the data and estimated for each voxel an ensemble of $\mathit{N}_\mathrm{b}=96$ plausible sets of components, also called "bootstrap solutions", each denoted by $\{(\mathit{D}_{\parallel,\mathit{n}}, \mathit{D}_{\perp,\mathit{n}} , \theta_\mathit{n}, \phi_\mathit{n}, \mathit{R}_{1,\mathit{n}}, \mathit{w}_\mathit{n})\}_{1\leq \mathit{n}\leq \mathit{N}=20}$. We then computed statistical descriptors of $\mathcal{P}(\mathbf{D},\mathit{R}_1)$ for each bootstrap solution and calculated the median of each statistical descriptor across bootstrap solutions (see Section~\ref{Sec_statistical_descriptors_binning}).


\subsubsection{Statistical descriptors and binning}
\label{Sec_statistical_descriptors_binning}

The final solution of the Monte-Carlo inversion algorithm, $\mathcal{P}(\mathbf{D},\mathit{R}_1)$, can be understood as the median of the solutions obtained for each bootstrap solution, $\mathcal{P}_{\mathit{n}_\mathrm{b}}(\mathbf{D},\mathit{R}_1)$, with $1\leq \mathit{n}_\mathrm{b} \leq \mathit{N}_\mathrm{b}=96$. Following previous works,~\citep{Reymbaut_accuracy_precision:2020,deAlmeidaMartins:2020} we quantified the main features of this final solution by computing the median across bootstrap solutions of means $\mathrm{Med}_{(n_\mathrm{b})}\,(\mathrm{E}[\chi]_{n_\mathrm{b}})$, variances $\mathrm{Med}_{(n_\mathrm{b})}\,(\mathrm{V}[\chi]_{n_\mathrm{b}})$ and covariances $\mathrm{Med}_{(n_\mathrm{b})}\,(\mathrm{C}[\chi,\chi^\prime]_{n_\mathrm{b}})$ of the per-bootstrap isotropic diffusivity, squared normalized anisotropy, and longitudinal relaxation rate $\chi,\chi^\prime=\mathit{D}_\mathrm{iso}, D_\Delta^2, R_1$, respectively. Here, the median operator $\mathrm{Med}_{(n_\mathrm{b})}(\,\cdot\,)$ acts across bootstrap solutions and $\mathrm{E}[\,\cdot\,]_{\mathit{n}_\mathrm{b}}$, $\mathrm{V}[\,\cdot\,]_{\mathit{n}_\mathrm{b}}$ and $\mathrm{C}[\,\cdot,\cdot\,]_{\mathit{n}_\mathrm{b}}$ denote the per-bootstrap average, variance and covariance over the diffusion-relaxation components forming the bootstrap solution $\mathit{n}_\mathrm{b}$, respectively. For simplicity, we omit the explicit mention of the median operator when addressing a statistical descriptor, thereby writing averages, variances and covariances as $\mathrm{E}[\chi]$, $\mathrm{V}[\chi]$ and $\mathrm{C}[\chi,\chi^\prime]$ respectively.

By design, the Monte-Carlo inversion algorithm progressively builds up the sought-for intra-voxel distribution $\mathcal{P}(\mathbf{D},\mathit{R}_1)$ as a non-parametric discrete weighted sum of components. This implies that tissue-specific statistical descriptors can be extracted by subdividing the 5D configuration space of $\mathcal{P}(\mathbf{D},\mathit{R}_1)$ into multiple bins. For instance, the "thin", "thick" and "big" bins introduced in Refs.~\onlinecite{Topgaard:2019, deAlmeidaMartins:2020} aim to isolate the signal contributions from white matter, grey matter and cerebrospinal fluid, respectively. In this work, the boundaries of these bins, illustrated in the panels C, D and E of Figure~\ref{Figure_fit_distributions}, were defined as follows: 
\begin{itemize}
\item "big" bin within $D_\mathrm{iso} \in [2, 10] \;\text{\textmu}\mathrm{m}^2/\mathrm{ms}$, $D_\parallel/D_\perp \in [0.01, 1000]$ and $R_1 \in [0.01, 10] \;\mathrm{s}^{-1}$.
\item "thick" bin within $D_\mathrm{iso} \in [0.1, 2] \;\text{\textmu}\mathrm{m}^2/\mathrm{ms}$, $D_\parallel/D_\perp \in [0.01, 4]$ and $R_1 \in [0.01, 10] \;\mathrm{s}^{-1}$.
\item "thin" bin within $D_\mathrm{iso} \in [0.1, 2] \;\text{\textmu}\mathrm{m}^2/\mathrm{ms}$, $D_\parallel/D_\perp \in [4, 1000]$ and $R_1 \in [0.01, 10] \;\mathrm{s}^{-1}$.
\end{itemize}
As such, the "big" bin captures highly diffusive components, the "thick" bin captures components that are not highly diffusive nor highly anisotropic, and the "thin" bin captures components that are highly anisotropic. Note that the above bin boundaries extend far beyond the inversion boundaries listed in Section~\ref{Sec_MC_algorithm}, so that to produce aesthetically pleasing figures such as the panels C, D and E of Figure~\ref{Figure_fit_distributions}. Bin-specific statistical descriptors were estimated following the aforementioned process for the retrieved components that specifically fall into each bin. 

\subsubsection{Orientation distribution functions (ODFs)}
\label{Sec_ODFs}

Orientation distribution functions (ODFs) can be defined from the thin-bin components output by the Monte-Carlo signal inversion of Section~\ref{Sec_MC_algorithm} using the procedure detailed in Refs.~\onlinecite{deAlmeidaMartins_ODF:2020, deAlmeidaMartins_thesis:2020}. Briefly, per-bootstrap ODFs were generated by convolving the discrete ensemble of weights $w_i$ and unit-orientations $(\theta_i, \phi_i)$ of the components belonging to the thin bin in each bootstrap solution $n_\mathrm{b}$, $\{(w_\mathit{i}, \theta_i, \phi_i)\}_{\mathit{n}_\mathrm{b},\;\mathit{i}\in \{\text{thin bin}\}}$, with a Watson kernel. The purpose of this kernel is to smoothly map the discrete set of components onto the nearest nodes of a dense spherical mesh $\{(\theta_{\text{mesh}},\phi_{\text{mesh}})\}$. A final voxel-wise ODF $\mathit{P}(\theta_{\text{mesh}},\phi_{\text{mesh}})$ was calculated as the median of the per-bootstrap ODFs. In this work, we considered a $1000$-point uniform spherical mesh and set the concentration parameter of the Watson kernel to $\kappa=14.9$, following the rationale detailed in Ref.~\onlinecite{deAlmeidaMartins_ODF:2020}.

Following a similar procedure, the discrete set of diffusion-relaxation metrics $\{(\mathit{D}_{\parallel,\mathit{i}}, \mathit{D}_{\perp,\mathit{i}}, \mathit{R}_{1,\mathit{i}})\}_{\mathit{n}_\mathrm{b},\;\mathit{i}\in \{\text{thin bin}\}}$ can also be mapped onto this spherical mesh,~\citep{deAlmeidaMartins_ODF:2020, deAlmeidaMartins_thesis:2020} yielding the orientation-specific diffusion-relaxation measures $\hat{\mathrm{E}}[\chi]_{\mathit{n}_\mathrm{b}}(\theta_\text{mesh},\phi_\text{mesh})$, with $\chi \equiv \mathit{D}_\mathrm{iso}, \mathit{D}_\Delta^2, \mathit{R}_1$. In addition to coloring ODFs according to the $(\theta_{\text{mesh}},\phi_{\text{mesh}})$ local orientation, this mapping allows to color ODFs according to the local value $\mathrm{Med}_{(n_\mathrm{b})}\left(\hat{\mathrm{E}}[\chi]_{\mathit{n}_\mathrm{b}}(\theta_\text{mesh},\phi_\text{mesh})\right)$, thereby improving the visualization of orientation-specific diffusion-relaxation quantities. For simplicity, the short-hand notation "$\hat{\mathrm{E}}[\chi]$" is now retained instead for $\mathrm{Med}_{(n_\mathrm{b})}\left(\hat{\mathrm{E}}[\chi]_{\mathit{n}_\mathrm{b}}(\theta_\text{mesh},\phi_\text{mesh})\right)$.

\subsection{Monte-Carlo density-peak clustering (MC-DPC)}
\label{Sec_MC_DPC}

The Monte-Carlo signal inversion algorithm of Section~\ref{Sec_MC_algorithm} can be combined with density-peak clustering (DPC)~\citep{Rodriguez_Laio:2014} according to the work presented in Ref.~\onlinecite{Reymbaut_arxiv_MC_DPC:2020}. This combination, called "Monte-Carlo density-peak clustering" (MC-DPC), enables to quantify the median value and precision of orientation-resolved means of $\chi=\mathit{D}_\mathrm{iso}, D_\Delta^2, R_1, T_1=1/R_1$ across bootstrap solutions. We used MC-DPC to detect statistically significant differences between sub-voxel fiber populations robustly assigned to major fiber bundles \textit{a posteriori}.

Firstly, MC-DPC gathers the ensemble $\{ \mathcal{E}^\mathrm{thin}_{n_\mathrm{b}} \}_{1\leq n_\mathrm{b}\leq N_\mathrm{b}}$ of all per-bootstrap thin-bin solution sets $\mathcal{E}^\mathrm{thin}_{n_\mathrm{b}} = \{(\mathit{D}_{\parallel,\mathit{i}}, \mathit{D}_{\perp,\mathit{i}} , \theta_\mathit{i}, \phi_\mathit{i}, \mathit{R}_{1,\mathit{i}}, \mathit{w}_\mathit{i})\}_{n_\mathrm{b},\;\mathit{i}\in \{\text{thin bin}\}}$ and delineates $N_\mathrm{c}$ clusters in its orientation subspace using DPC with data-point density and outlier detection altered to account for the weights $w_i$ of the retrieved thin-bin components.~\citep{Reymbaut_arxiv_MC_DPC:2020} An initial number of clusters $N_\mathrm{c}$ is automatically set by the number of voxel-wise ODF peaks, but may be reduced by MC-DPC following a filtering approach detailed in Ref.~\onlinecite{Reymbaut_arxiv_MC_DPC:2020}. Assuming that the estimated clusters, resulting from orientational aggregates of the all-bootstrap thin-bin solutions, can be interpreted as orientational regions of interest associated with sub-voxel fiber populations, MC-DPC then computes orientation-resolved statistics across bootstrap solutions. To do so, it separately classifies each per-bootstrap ensemble of thin-bin solutions $\mathcal{E}^\mathrm{thin}_{n_\mathrm{b}}$ into $N_\mathrm{c}$ ensembles $\mathcal{E}^\mathrm{thin}_{n_\mathrm{b},n_\mathrm{c}}$ (with $1\leq n_\mathrm{c}\leq N_\mathrm{c}$), each containing the thin-bin solutions of bootstrap solution $n_\mathrm{b}$ that belong to an estimated cluster $n_\mathrm{c}$. It then averages the properties of the solutions within each ensemble $\mathcal{E}^\mathrm{thin}_{n_\mathrm{b},n_\mathrm{c}}$ independently, yielding the orientation-resolved means
\begin{equation}
\mathring{\mathrm{E}}[\chi]_{n_\mathrm{b},n_\mathrm{c}} = \frac{\sum_{k\in  \mathcal{E}^\mathrm{thin}_{n_\mathrm{b},n_\mathrm{c}}} w_k\,\chi_k}{\sum_{k\in \mathcal{E}^\mathrm{thin}_{n_\mathrm{b},n_\mathrm{c}}} w_k} \, ,
\label{Eq_orientation_resolved_mean}
\end{equation}
with $\chi\equiv x,y,z, \mathit{D}_\mathrm{iso}, \mathit{D}_\Delta^2, \mathit{R}_1, \mathit{T}_1=1/\mathit{R}_1$, where $(x,y,z)$ are the Cartesian coordinates of a component's orientation $(\theta,\phi)$. The short-hand notation "$\mathring{\mathrm{E}}[\chi]$" will be used for simplicity to describe the collection of orientation-resolved means $\mathring{\mathrm{E}}[\chi]_{n_\mathrm{b},n_\mathrm{c}}$ originating from all bootstrap solutions $n_\mathrm{b}$ and all clusters $n_\mathrm{c}$. The orientation-resolved means of $T_1$ and $R_1$ in Equation~\ref{Eq_orientation_resolved_mean} are computed separately, as both quantities are commonly found in the MRI literature and $\mathring{\mathrm{E}}[T_1]$ does not equal $1/\mathring{\mathrm{E}}[R_1]$. Finally, we extracted the median and interquartile range of our various orientation-resolved means $\mathring{\mathrm{E}}[\chi]$ across bootstrap solutions.

Non-parametric Mann-Whitney $U$-tests were used to assess whether or not two orientation-resolved means $\mathring{\mathrm{E}}[\chi]$ characterizing distinct white-matter fiber bundles are sampled from identically shaped non-median-shifted continuous distributions (null hypothesis $\mathcal{H}_0$). The $p$-values resulting from these tests inform on the acceptance or rejection of $\mathcal{H}_0$ at a certain significance level. In particular, rejections of $\mathcal{H}_0$ at $p < 0.01$, $0.01\leq p < 0.05$ and $0.05\leq p < 0.1$ were used as proxies for detecting significant differences between two bundles with respect to a given type of orientation-resolved mean.

\section{Results}
\label{Sec_Results}

\begin{figure*}[ht!]
\begin{center}
\includegraphics[width=30pc]{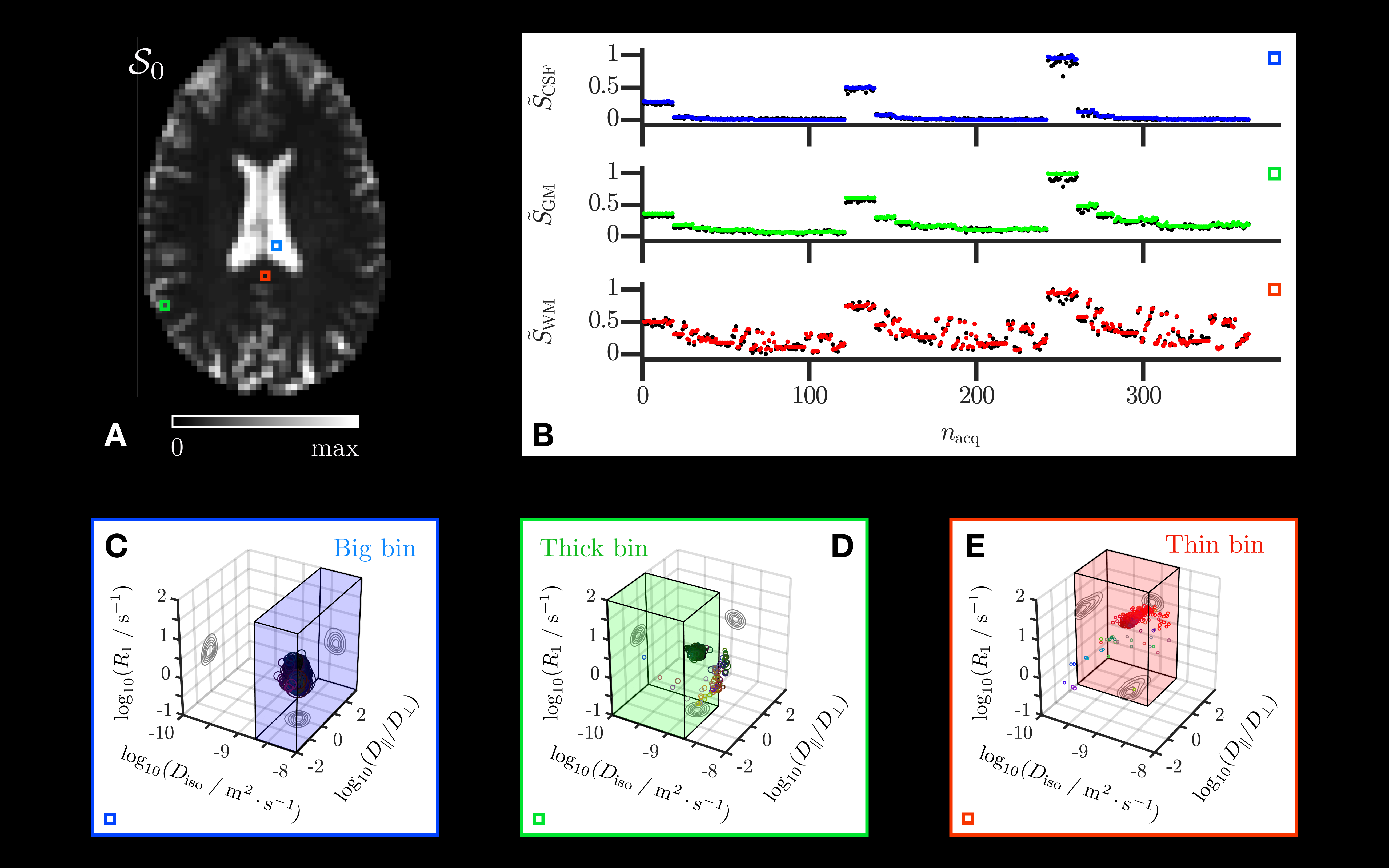}
\caption{Monte-Carlo fitted signal and retrieved 5D distributions $\mathcal{P}(\mathbf{D},R_1)$ in typical voxels. (A) $\mathcal{S}_0$ map estimated by the Monte-Carlo inversion. The colored squares delineate typical CSF (blue), GM (green) and WM (red) voxels. (B) Normalized signal $\tilde{S} = \mathcal{S}/\mathrm{max}(\mathcal{S})$ measured (black points) and fitted (colored points) in the archetypal voxels of panel A as a function of the sorted acquisition point index $n_\mathrm{acq}$ of Figure~\ref{Figure_acq}. (C,D,E) Non-parametric distributions $\mathcal{P}(\mathbf{D},R_1)$ estimated within the archetypal voxels of panel A and reported as scatter plots in a 3D space of the logarithms of the longitudinal relaxation rate $R_1$, isotropic diffusivity $D_\mathrm{iso}$, and axial-radial diffusivity ratio $D_\parallel/D_\perp$. Diffusion-orientations $(\theta, \phi)$ are color-coded according to $[\mathrm{red,green,blue}] = [\sin\theta\cos\phi, \sin\theta\sin\phi, \cos\theta]\times \vert D_\parallel - D_\perp\vert/\mathrm{max}(D_\parallel,D_\perp)$. Symbol area is proportional to the statistical weight $w_n/\mathcal{S}_0$ of the corresponding component $n$. The contour lines on the sides of the plots represent projections of the 5D distributions $\mathcal{P}(\mathbf{D},R_1)$ onto the respective 2D planes. The "thin", "thick" and "big" bins defined in Section~\ref{Sec_statistical_descriptors_binning} are illustrated in the panels where they are most relevant.}
\label{Figure_fit_distributions}
\end{center}
\end{figure*}

\begin{figure*}[ht!]
\begin{center}
\includegraphics[width=30pc]{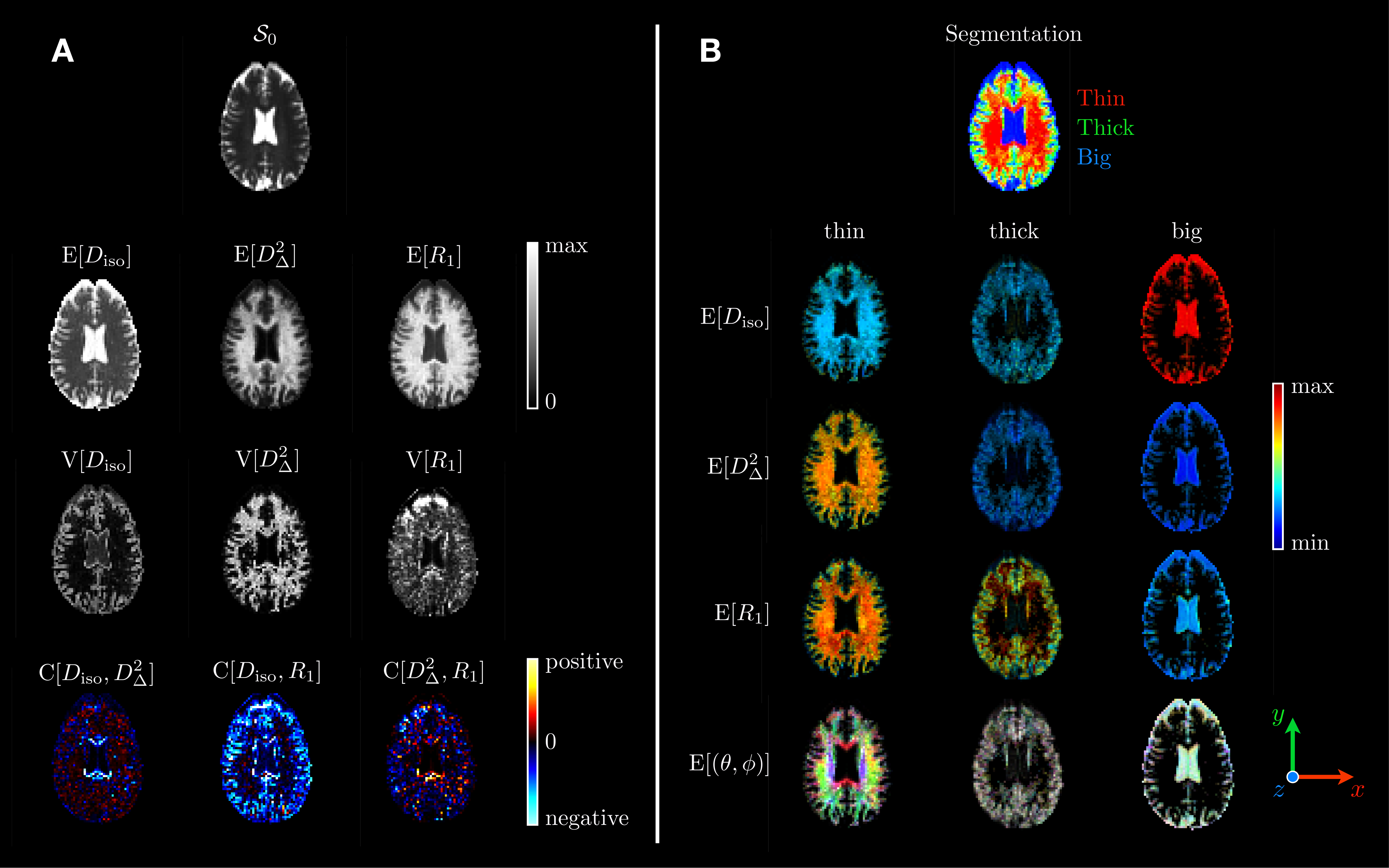}
\caption{Typical axial maps of the statistical descriptors described in Section~\ref{Sec_statistical_descriptors_binning}. The various colormaps are bounded by the following values: $\mathcal{S}_0\in[0, \mathrm{max}(\mathcal{S}_0)]$, $\mathrm{E}[D_\mathrm{iso}]\in[0, 3.5]\;\text{\textmu}\mathrm{m}^2/\mathrm{ms}$, $\mathrm{E}[D_\Delta^2]\in[0, 1]$, $\mathrm{E}[R_1]\in [0,0.8]\;\mathrm{s}^{-1}$, $\mathrm{V}[D_\mathrm{iso}]\in[0, 0.9]\;\text{\textmu}\mathrm{m}^4/\mathrm{ms}^2$, $\mathrm{V}[D_\Delta^2]\in[0, 0.1]$, $\mathrm{V}[R_1]\in[0, 0.13]\;\mathrm{s}^{-2}$, $\mathrm{C}[D_\mathrm{iso},D_\Delta^2]\in[-0.3, 0.3]\;\text{\textmu}\mathrm{m}^2/\mathrm{ms}$, $\mathrm{C}[D_\mathrm{iso},R_1]\in[-2.4, 2.4]\times 10^{-4}\;\text{\textmu}\mathrm{m}^2/\mathrm{ms}^2$ and $\mathrm{C}[D_\Delta^2,R_1]\in[-0.08, 0.08]\;\mathrm{s}^{-1}$. The bin-specific average intra-voxel orientation $\mathrm{E}[(\theta,\phi)]$ is color-coded for orientation according to $[\mathrm{red,green,blue}] = [\mathrm{E}[D_{xx}], \mathrm{E}[D_{yy}], \mathrm{E}[D_{zz}]]/\mathrm{max}(\mathrm{E}[D_{xx}], \mathrm{E}[D_{yy}], \mathrm{E}[D_{zz}])$, where the average diffusivities $\mathrm{E}[D_{ii}]$ are associated with the directions $i=x, y, z$ corresponding to the "left-right", "anterior-posterior" and "superior-inferior" directions, respectively. For a given bin, the intensity of the bin-specific maps equals the voxel-wise average fraction $f_\mathrm{bin}$ of components belonging to this bin. The segmentation map is colored according to $[\mathrm{red,green,blue}] = [f_\mathrm{thin}, f_\mathrm{thick}, f_\mathrm{big}]/\mathrm{max}(f_\mathrm{thin}, f_\mathrm{thick}, f_\mathrm{big})$.}
\label{Figure_maps}
\end{center}
\end{figure*}

\begin{figure*}[ht!]
\begin{center}
\includegraphics[width=30pc]{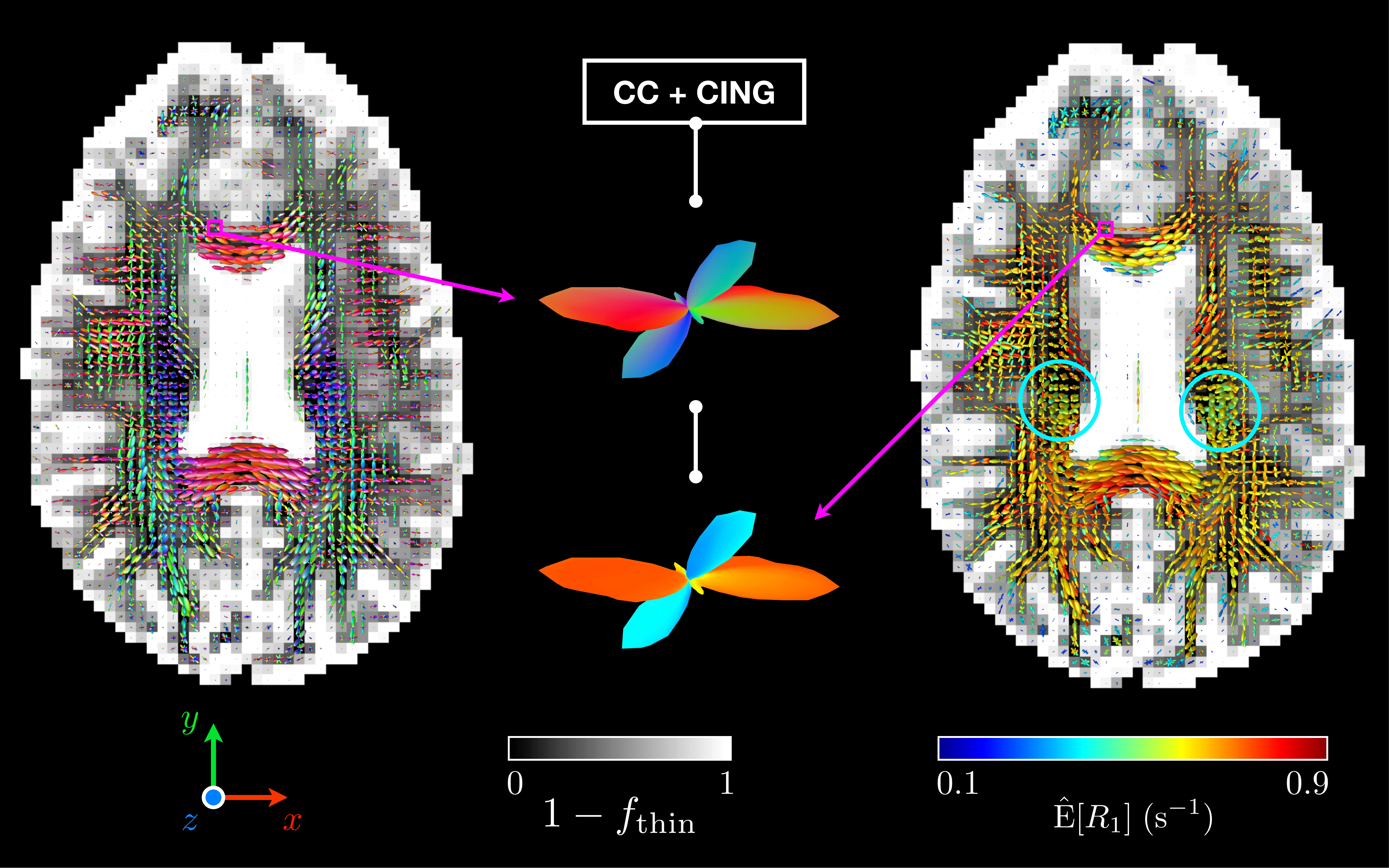}
\caption{Axial greyscale maps of the fraction of non thin-bin components $1-f_\mathrm{thin}$ with superimposed ODFs colored by local orientation (with $x$, $y$ and $z$ corresponding to the "left-right", "anterior-posterior" and "superior-inferior" directions, respectively) and by $\hat{\mathrm{E}}[R_1]$ (see Section~\ref{Sec_ODFs}). The middle insets zoom on a voxel containing a fiber crossing between the corpus callosum (CC) and the cingulum (CING), and presents the estimated orientation-colored and $\hat{\mathrm{E}}[R_1]$-colored ODFs for this voxel. While differences in $\hat{\mathrm{E}}[R_1]$ seem to exist between CC and CING, such differences may also exist in the regions where the CST's pyramidal tracts are located (blue circles), as indicated by the greener $\hat{\mathrm{E}}[R_1]$-colored ODFs therein.}
\label{Figure_ODFs}
\end{center}
\end{figure*}

\begin{figure*}[ht!]
\begin{center}
\includegraphics[width=33pc]{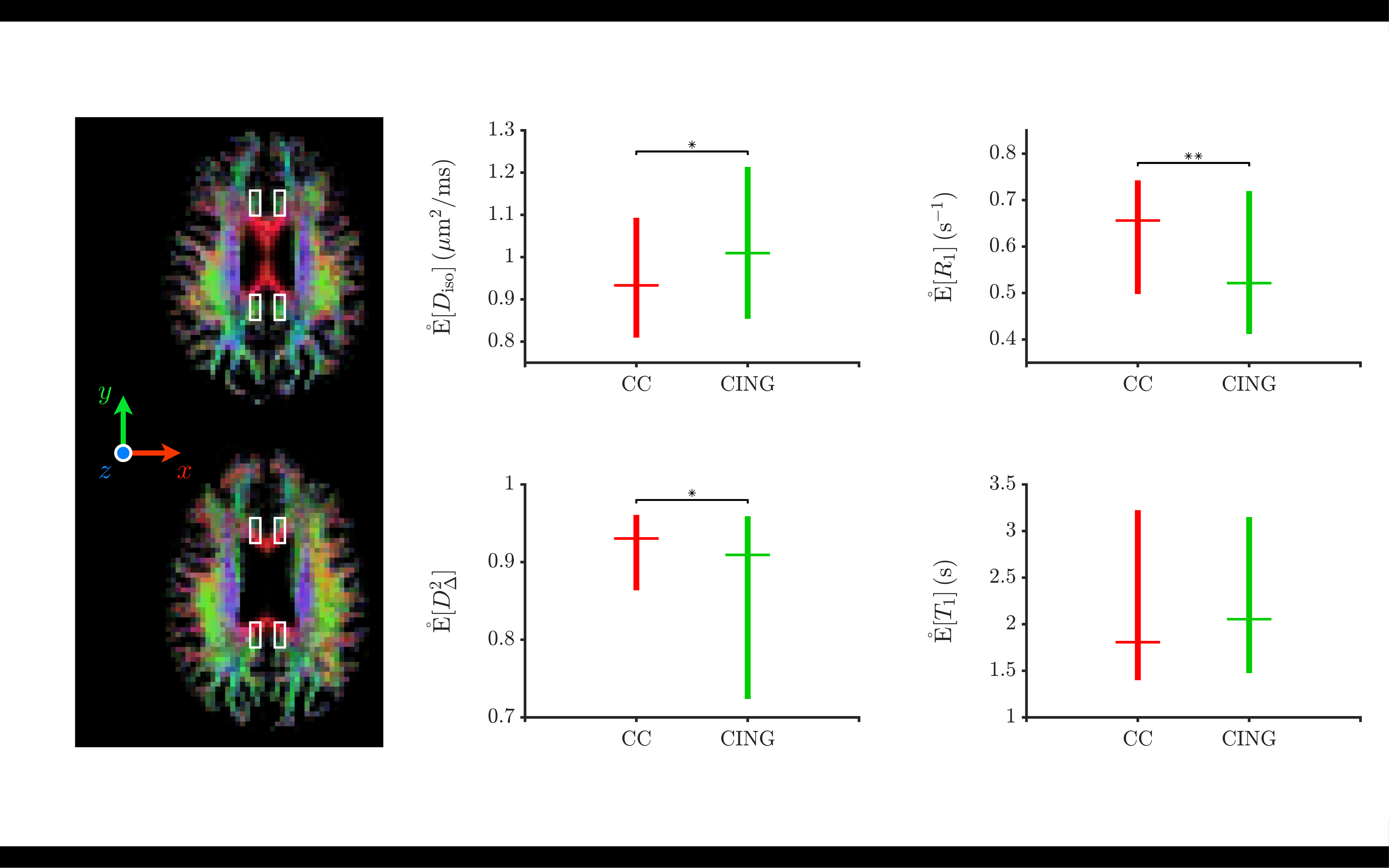}
\caption{Boxplots of the medians of the orientation-resolved means $\mathring{\mathrm{E}}[\chi]$ obtained from MC-DPC (see Equation~\ref{Eq_orientation_resolved_mean}) within the regions of interest drawn in the left panel (white-lined boxes), showing two axial slices of the orientation-colored average fraction of thin-bin components $f_\mathrm{thin}$. For a given boxplot, the horizontal line and whiskers indicate the median and the range between the first and third quartiles of the medians of the orientation-resolved means $\mathring{\mathrm{E}}[\chi]$, respectively. The chosen regions of interest focus on crossing areas between the corpus callosum (CC) and the cingulum (CING). Each MC-DPC cluster (and associated orientation-resolved means) is robustly assigned to one of these bundles depending on whether its median orientation is closer to the $x$ "left-right" direction (CC) or to the $y$ "anterior-posterior" direction (CING). The asterisks report the results of non-parametric Mann-Whitney $U$-tests assessing whether or not two orientation-resolved means $\mathring{\mathrm{E}}[\chi]$ assigned to distinct bundles are sampled from identically shaped non-median-shifted continuous distributions (null hypothesis $\mathcal{H}_0$). The $p$-values resulting from these tests inform on the acceptance or rejection of $\mathcal{H}_0$ at a certain significance level: $0.05\leq p < 0.1$ (*) and $0.01\leq p < 0.05$ (**).}
\label{Figure_CC_CING}
\end{center}
\end{figure*}

\begin{figure*}[ht!]
\begin{center}
\includegraphics[width=33pc]{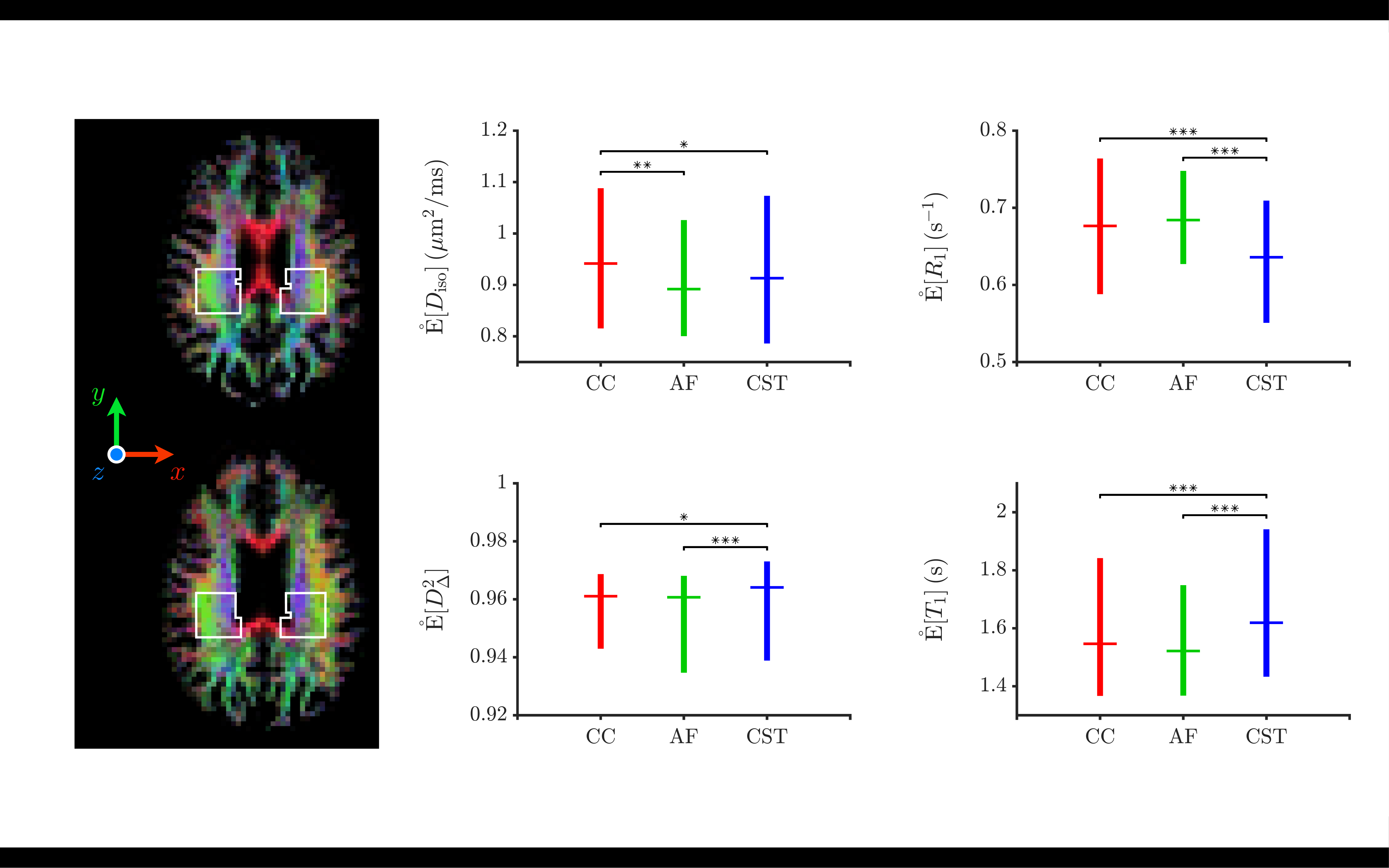}
\caption{Equivalent of Figure~\ref{Figure_CC_CING} for regions of interest focusing on crossing areas between the corpus callosum (CC), the arcuate fasciculus (AF) and the corticospinal tract (CST) in the posterior corona radiata. Each MC-DPC cluster (and associated orientation-resolved means) is robustly assigned to one of these bundles depending on whether its median orientation is closer to the $x$ "left-right" direction (CC), to the $y$ "anterior-posterior" direction (AF), or to the $z$ "superior-inferior" direction (CST). The asterisks report the results of non-parametric Mann-Whitney $U$-tests assessing whether or not two orientation-resolved means $\mathring{\mathrm{E}}[\chi]$ assigned to distinct bundles are sampled from identically shaped non-median-shifted continuous distributions (null hypothesis $\mathcal{H}_0$). The $p$-values resulting from these tests inform on the acceptance or rejection of $\mathcal{H}_0$ at a certain significance level: $0.05\leq p < 0.1$ (*), $0.01\leq p < 0.05$ (**) and $p < 0.01$ (***).}
\label{Figure_CS}
\end{center}
\end{figure*}

Figure~\ref{Figure_fit_distributions} presents the fitted signals and the intra-voxel distributions $\mathcal{P}(\mathbf{D},R_1)$ estimated by the Monte-Carlo signal inversion algorithm of Section~\ref{Sec_MC_algorithm} in typical voxels associated with cerebrospinal fluid (CSF) in the ventricles, cortical grey matter (GM) and white matter (WM) in the corpus callosum. Figure~\ref{Figure_maps} displays typical axial maps of the statistical descriptors and bin-specific statistical descriptors described in Section~\ref{Sec_statistical_descriptors_binning}. Figure~\ref{Figure_ODFs} shows orientation-colored and $\hat{\mathrm{E}}[R_1]$-colored ODFs (see Section~\ref{Sec_ODFs}) in a typical axial slice. Figures~\ref{Figure_CC_CING} and \ref{Figure_CS} investigate possible microstructural differences between sub-voxel fiber populations leveraging MC-DPC (see Section~\ref{Sec_MC_DPC}) in regions of interest that target specific fiber crossings. While Figure~\ref{Figure_CC_CING} focuses on the crossing between the corpus callosum (CC) and the cingulum (CING), Figure~\ref{Figure_CS} focuses on the crossing between the corpus callosum (CC), the arcuate fasciculus (AF) and the corticospinal tract (CST) in the posterior corona radiata.

\section{Discussion}
\label{Sec_Discussion}

Figure~\ref{Figure_fit_distributions} demonstrates that the Monte-Carlo signal inversion algorithm described in Section~\ref{Sec_MC_algorithm} adequately fits the measured signal in various voxels pertaining to WM, cortical GM and CSF in the ventricles. Additionally, the bins defined in Section~\ref{Sec_statistical_descriptors_binning} appear to capture these environments in accordance with their original design, \textit{i.e.} the thin, thick and big bins capture components typically associated with WM, GM and CSF, respectively.

Figure~\ref{Figure_maps} shows that the Monte-Carlo signal inversion algorithm can estimate maps of $\mathcal{P}(\mathbf{D},R_1)$'s statistical descriptors. In particular, it retrieves maps that are consistent with those thoroughly discussed in Ref.~\onlinecite{deAlmeidaMartins:2020}, namely the $\mathcal{S}_0$, $\mathrm{E}[D_\mathrm{iso}]$, $\mathrm{E}[D_\Delta^2]$, $\mathrm{V}[D_\mathrm{iso}]$, $\mathrm{V}[D_\Delta^2]$ and $\mathrm{C}[D_\mathrm{iso},D_\Delta^2]$ maps, along with the bin-specific $\mathrm{E}[D_\mathrm{iso}]$, $\mathrm{E}[D_\Delta^2]$ and average local orientation $\mathrm{E}[(\theta,\phi)]$ maps, and the bin-segmentation map. Let us thus mainly discuss the $R_1$-related maps. The $\mathrm{E}[R_1]$ map resembles an expected low-resolution $T_1$ map, \textit{i.e.} bright in WM, slightly darker in GM and very dark in CSF. The $\mathrm{V}[R_1]$ map resembles a noisier version of the $\mathrm{V}[D_\mathrm{iso}]$ map, as both give high values in mixed CSF-WM/GM voxels. The noise in $\mathrm{V}[R_1]$ could be reduced upon adding more repetition times in the acquisition scheme. The $\mathrm{C}[D_\mathrm{iso},R_1]$ map is negative at the interface between CSF and either WM or cortical GM. Indeed, upon entering CSF from WM/GM, $D_\mathrm{iso}$ increases rapidly and $T_1$ relaxation slows down (rapidly decreasing $R_1$). Finally, the $\mathrm{C}[D_\Delta^2,R_1]$ map exhibits no specific pattern. As for the bin-specific $\mathrm{E}[R_1]$ maps, they present a clear contrast between our tissue-specific bins, due to a fast $T_1$ relaxation in WM, a slower relaxation in GM and an even slower relaxation in CSF.

Figure~\ref{Figure_ODFs} features non-parametric ODFs capturing local orientations that are consistent with the known anatomy. Regarding $\hat{\mathrm{E}}[R_1]$-colored ODFs (see Section~\ref{Sec_ODFs}), they appear to change colors when approaching tissue interfaces with CSF. This gradual change in $\hat{\mathrm{E}}[R_1]$ may originate from exchange between tissues and CSF in these regions. Importantly, Figure~\ref{Figure_ODFs} shows that potential differences in $T_1$ relaxation may exist between major fiber bundles, namely CC and CING (insets in Figure~\ref{Figure_ODFs}), and CC, AF and CST (as indicated by greener $\hat{\mathrm{E}}[R_1]$-colored ODFs where the CST's pyramidal tracts are located).

These potential microstructural differences are quantified in Figures~\ref{Figure_CC_CING} and \ref{Figure_CS}. Let us focus on relaxation-based differences. Figure~\ref{Figure_CC_CING} shows that CC and CING exhibit significant differences in $\mathring{\mathrm{E}}[R_1]$ that are qualitatively consistent with those found in Refs.~\onlinecite{deSantis_T1:2016, Andrews_ISMRM:2019}, \textit{i.e.} $R_1$ tends to be lower in CING compared to CC. As for Figure~\ref{Figure_CS}, it shows that CST features significant differences in $\mathring{\mathrm{E}}[R_1]$ and $\mathring{\mathrm{E}}[T_1]$ with CC and AF (but no statistically significant differences between CC and AF). These differences are qualitatively consistent with those identified for CST in Ref.~\onlinecite{deSantis_T1:2016}, \textit{i.e.} $T_1$ tends to be higher in CST compared to CC and AF. 

Quantitatively, the $T_1$ values estimated by $\mathring{\mathrm{E}}[T_1]$ in Figures~\ref{Figure_CC_CING} and \ref{Figure_CS} (around 1.5 to 2 seconds) are overestimated compared to those of Ref.~\onlinecite{deSantis_T1:2016} (around 0.9 to 1 second) and Ref.~\onlinecite{Andrews_ISMRM:2019} (around 0.7 second). This discrepancy can be explained by either or both of the following factors. Firstly, the acquisition scheme described in Section~\ref{Sec_in_vivo_data} does not does not maximize the amount of diffusion-relaxation correlations built into the inversion kernel of Equation~\ref{Eq_signal_discretized}, because the same diffusion-weighting block was repeated for each acquired repetition time. Similar problems have been suggested to lead to a loss of accuracy for the Monte-Carlo signal inversion.~\citep{Reymbaut_accuracy_precision:2020} Second, the use of saturation recovery with a spoiled spin echo for $T_1$ encoding is very sensitive to flip-angle inaccuracies caused by both $B_1^+$ inhomogeneity across the subject and slice profile imperfections. Saturation-recovery based $T_1$ mapping is also sensitive to magnetization-transfer effects, especially in the present setup comprising an additional refocusing pulse as well as a fat-saturation pulse.~\citep{Wolff_Balaban:1989, Teixeira:2019, Teixeira:2020} These technical limitations should be mitigated upon developing a sequence that includes inversion preparation for enhanced $T_1$ sensitivity and slice shuffling for optimized time efficiency.~\citep{Hutter:2018, Park_ISMRM:2018}

\newpage

\section{Conclusions}
\label{Sec_Conclusions}

Diffusion-$T_1$ weighted datasets incorporating multiple b-tensor shapes can be inverted to obtain non-parametric distributions $\mathcal{P}(\mathbf{D},R_1)$ of diffusion tensors and longitudinal relaxation rates using the Monte-Carlo signal inversion algorithm. The main features of the retrieved distributions can be visualized as maps and bin-specific maps of statistical descriptors related to means, variances and covariances of diffusion-relaxation properties. In particular, the bin-specific $\mathrm{E}[R_1]$ maps exhibit the expected $R_1$ contrast between white matter, grey matter and CSF. Further insight into white-matter microstructure is provided by the "thin bin", which isolates highly anisotropic components that should report on white-matter tissues. From these thin-bin components, visualization of fiber-specific information is improved upon defining orientation distribution functions (ODFs) that can be color-mapped with respect to local orientation or diffusion-relaxation features.~\citep{deAlmeidaMartins_ODF:2020} While $\hat{\mathrm{E}}[R_1]$-colored ODFs hint at possible differences between fiber bundles, Monte-Carlo density-peak clustering (MC-DPC) enables their quantification in terms of fiber-specific diffusion-relaxation measures.~\citep{Reymbaut_arxiv_MC_DPC:2020}

Importantly, significant differences with respect to longitudinal relaxation are detected between the corpus callosum and the cingulum, and between the corticospinal tract and the corpus callosum and arcuate fasciculus. These differences, qualitatively consistent with those found in previous works,~\citep{deSantis_T1:2016, Andrews_ISMRM:2019} offer a proof of concept for the potential of our Monte-Carlo framework in terms of non-parametric fiber-specific $T_1$ relaxometry. Such approach would be practical in identifying differences in $\mathit{T}_1$ between distinct sub-voxel fiber populations, characterizing developmental or pathological changes in $\mathit{T}_1$ within a given sub-voxel fiber population, and measuring the angular dependence of longitudinal relaxation times in white matter with respect to the main MRI magnetic field $\mathbf{B}_0$.~\citep{Henkelman:1994,Knight:2018} Moreover, non-parametric fiber-specific $T_1$ relaxometry would be highly relevant to microstructure-informed tractography.~\citep{Daducci:2016, Girard:2017, Barakovic_thesis:2019} 

Nevertheless, our work can still be improved in two main ways. Firstly, the acquisition scheme could be optimized in terms of speed~\citep{Hutter:2018, Park_ISMRM:2018} and sensitivity.~\citep{Song:2005, Bates_ISMRM:2019, Song:2020} Second, MC-DPC could be combined with tractography to better assign MC-DPC's output clusters to their corresponding fiber bundles. These ideas will be explored in future investigations.

\newpage
\clearpage
\newpage

\section*{Acknowledgments}
This work was financially supported by the Swedish Foundation for Strategic Research (ITM17-0267) and the Swedish Research Council (2018-03697). D. Topgaard owns shares in Random Walk Imaging AB (Lund, Sweden, \href{http://www.rwi.se/}{http://www.rwi.se/}), holding patents related to the described methods.


%

\end{document}